\documentclass[a4paper,twoside]{article}
\usepackage{AMS2LA,fancyhdr,CJK,multicol,graphics,graphicx,bm,ccmap}
\usepackage[pdfstartview=FitH, bookmarks=false,colorlinks=false,
        linkcolor=blue, urlcolor=blue, citecolor=blue,
        pdftitle={Changepdftitle}, pdfauthor={Changepdfauthor},
        pdfcreator=CHIN.\ PHYS.\ LETT., pdfproducer=CPL,
        pdfkeywords={changePACS}]{hyperref}
\def\headrule{\kern 0mm \hrule width 0cm \kern -1mm}%
\def\footnoterule{\kern 1mm \hrule width 7cm \kern 2.2mm}%
\def\REF#1{\par\hangindent\parindent\indent\llap{#1\enspace}\ignorespaces}%
\begin{document}
\begin{CJK}{GBK}{song}\vspace* {-6mm}
\begin{center}

%\makeatletter
%\def\@cite#1#2{\textsuperscript{[{#1\if@tempswa , #2\fi}]}}
%\makeatother

%---------------------------Title --------------------------------
\large\bf{\boldmath{Possible Nodeless Superconducting Gaps in Bi$_2$Sr$_2$CaCu$_2$O$_{8+\delta}$ and YBa$_2$Cu$_3$O$_{7-x}$ Revealed by Cross-Sectional Scanning Tunneling Spectroscopy}}  
%\\[1mm]
%--------------------------Footnote--------------------------
\footnote{Supported in part by the National Science Foundation of China, and the National Key Research and Development Program of China (Grant No.\ 2016YFA0300203). We are grateful to Xin Yao for providing YBCO thin film samples, Hiroshi Eisaki for providing Bi2212 single crystals, and  Jiangping Hu, Yan Chen,  Tao Xiang, Ruihua He, Darren Peets and Yihua Wang for helpful discussions.

\hspace*{1.8mm}$^{**}$Correspondence author. Email:tzhang18@fudan.edu.cn

\hspace*{1.8mm}$^{***}$Correspondence author. Email:dlfeng@fudan.edu.cn }
%Tel: 021-51630266}
%\hspace*{1.8mm}\copyright\,{\cplyear}
%\href{http://www.cps-net.org.cn}{Chinese Physical Society} and
%\href{http://www.iop.org}{IOP Publishing Ltd}}
%\\[3mm]
%---------------------------Authors---------------------------
\\[4mm]
\normalsize \rm{} M. Q. Ren $^{1,2}$, Y. J. Yan $^{1,2}$, T. Zhang $^{1,2**}$ and D. L. Feng $^{1,2***}$
%--------------------------COM. or University -------------------
\\[2mm]
\small\sl $^{1}$ State Key Laboratory of Surface Physics, Department of Physics, and Advanced Materials Laboratory, Fudan University, Shanghai 200433, People's Republic of China
\\
\small\sl $^2$Collaborative Innovation Center of Advanced Microstructures, Nanjing 210093, People's Republic of China
\\[5mm]
%------------------------ Received date----------------------
%\normalsize\rm{}(Received 8 November 2016)

\end{center}
\end{CJK}
%----------------------Abstract and PACS---------------------
\vskip -1mm
%\maketitle

\noindent{\narrower\small\sl{}Pairing in the cuprate high-temperature superconductors and its origin remain among the most enduring mysteries in condensed matter physics.  
With cross-sectional scanning tunneling microscopy/ spectroscopy, we clearly reveal the spatial-dependence or inhomogeneity of the
superconducting gap structure of Bi$_2$Sr$_2$CaCu$_2$O$_{8+\delta}$ (Bi2212) and YBa$_2$Cu$_3$O$_{7-x}$ (YBCO) along their $c$-axes on a scale shorter than the interlayer spacing.
By tunneling into the (100) plane of a Bi2212 single crystal and a YBCO film, we observe both U-shaped tunneling spectra with extended flat zero-conductance bottoms, and V-shaped gap structures, in different regions of each sample.
On the YBCO film, tunneling into a (110) surface only reveals a U-shaped gap without any zero-bias peak.  Our analysis suggests that the U-shaped gap is likely a nodeless superconducting gap. The V-shaped gap has a very small amplitude, and is likely proximity-induced by regions having the larger U-shaped gap.
\par}\vskip 3mm\normalsize

\noindent{\narrower\sl{PACS:74.20.Rp;74.25.Jb;74.55.+v;74.72.-h. }
%{\rm\hspace*{13mm}	
%DOI:10.1000/0000-0000/\cplvol/\cplno/\cplpagenumber}

\par}\vskip 3mm
%--------------------TEXT TEXT TEXT TEXT---------------------
\begin{multicols}{2}

Cuprate high temperature superconductors are commonly understood to be composed of conducting CuO$_2$ layers and insulating charge reservoir oxide layers,$^{[1,2]}$ although, for example, early density functional theory calculations indicated metallic BiO layers in Bi$_2$Sr$_2$CaCu$_2$O$_{8+\delta}$ (Bi2212).$^{[3,4]}$ 
As a result, various transport properties and the low energy electronic structures of cuprates, as measured by angle-resolved photoemission spectroscopy (ARPES) and scanning tunneling microscopy/spectroscopy (STM/STS) are attributed solely to the CuO$_2$ planes.$^{[5-9]}$ In particular, the nodal superconducting gap and the pseudogap are thus believed to be critical characteristics of the CuO$_2$ planes. 

STS has played a critical role in our current understanding of the cuprates. For example, it has revealed strong in-plane inhomogeneity.$^{[8]}$ The $d$-wave nodal superconducting gap is manifested in the V-shaped local density of state (DOS) around zero bias with coherence peaks at the gap edges, while the pseudogap appears as a spectral weight suppression over hundreds of meV.$^{[7,9]}$ A competition between the superconducting gap and pseudogap features has been demonstrated in Bi2212,$^{[9]}$ and the latter was recently associated with states with broken spatial symmetries.

%%%%%%%%%% Fig. 1 %%%%%%%%%%%
%\vskip 0.5\baselineskip
%\vskip 4mm
\begin{figure*}[t]
%\centerline{
\includegraphics[width=\textwidth]{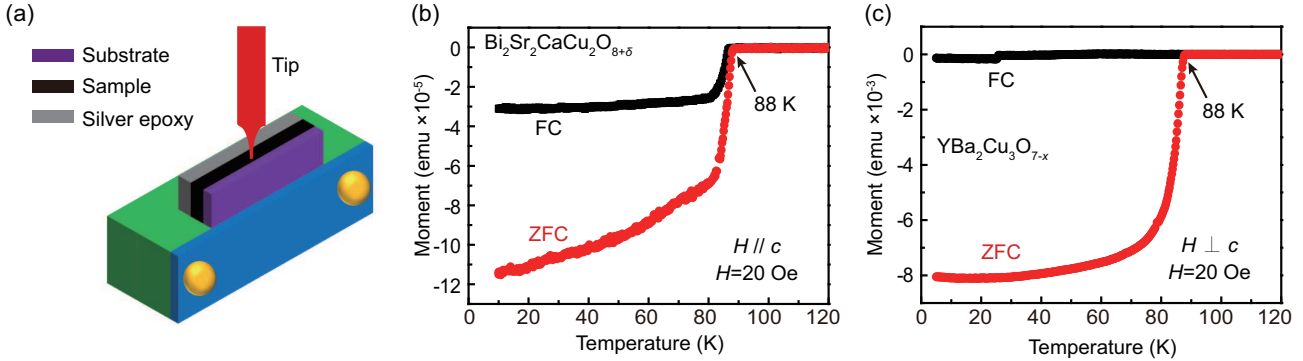}
%}
\caption{\label{fig1}(Color online).  (a) A sketch of the sample configuration. (b) and (c) Field-cooled (FC) and zero-field-cooled (ZFC) magnetic moment of Bi2212 single crystal and the YBCO film measured in a field of 20~Oe, respectively, which indicate that they both have a $T_c$ of 88 K. }
\end{figure*}
%\medskip
%%%%%%%%%% Fig. 1 %%%%%%%%%%%

However, this picture has been recently challenged by the STM study of CuO$_2$ planes  exposed by ion-sputtering or grown directly on the BiO surface of Bi2212.$^{[10-12]}$ A U-shaped gap structure is observed in $dI/dV$ spectra with an extended zero conductance region, indicating a nodeless superconducting gap. A V-shaped DOS is, however, observed on thick BiO layers (up to 4~nm) grown on Bi2212. These results hint that previous STM and ARPES measurements of the low-energy electronic structure near the Fermi energy ($E_F$) may contain contributions from both the CuO$_2$ layers and the charge reservoir layers, and those critical features, even  the $d$-wave gap itself, may {\slshape not} be intrinsic to the CuO$_2$ planes. 
On the other hand, the CuO$_2$ and BiO layers in these experiments are not in the usual bulk environment of Bi2212, since their doping levels and chemical environments, such as apical oxygen, are certainly different from the bulk. Therefore, it is easy to dismiss the observed phenomena as specific to these films and irrelevant to the situation of the bulk material or other cuprates. Moreover, a similar U-shaped tunneling spectrum has been found on a cleaved CuO$_2$ surface previously $^{[13]}$, and it could be fitted by a $d$-wave gap function after considering the fact that the c-axis tunneling is dominated by the $(\pi,0)$ region of the Brillouin zone, which is the antinodal region with the maximal gap. Therefore,  a U-shaped gap may not necessarily be inconsistent with the $d$-wave pairing.

To address these issues, we have conducted cross-sectional STS measurements of both Bi2212 and YBa$_2$Cu$_3$O$_{7-x}$ (YBCO), and achieved tunneling into side surfaces, including both (100) and (110) planes. These STS spectra directly access the electronic states of the CuO$_2$ plane and other layers, in a much closer approximation to the bulk environment. 
We uncover a strong spatial variation of the STS lineshape along the $c$-axis of the crystal, which enables the disentanglement of the highly layer-dependent low energy electronic structures in these compounds.
Remarkably, a clean U-shape particle-hole symmetric gap with an extended flat zero-conductance bottom is revealed without in-gap states at zero bias, and it could evolve into a smaller V-shaped gap with finite DOS at $E_F$ on a spatial scale of several nanometers. The observed gap structures provide new clues for studying the pairing symmetry of the cuprate superconductors.

%%%%%%%%%% Fig. 2 %%%%%%%%%%%
%\vskip 0.5\baselineskip
%\vskip 4mm
 \begin{figure*}[t!]
%\centerline{
\includegraphics[width=\textwidth]{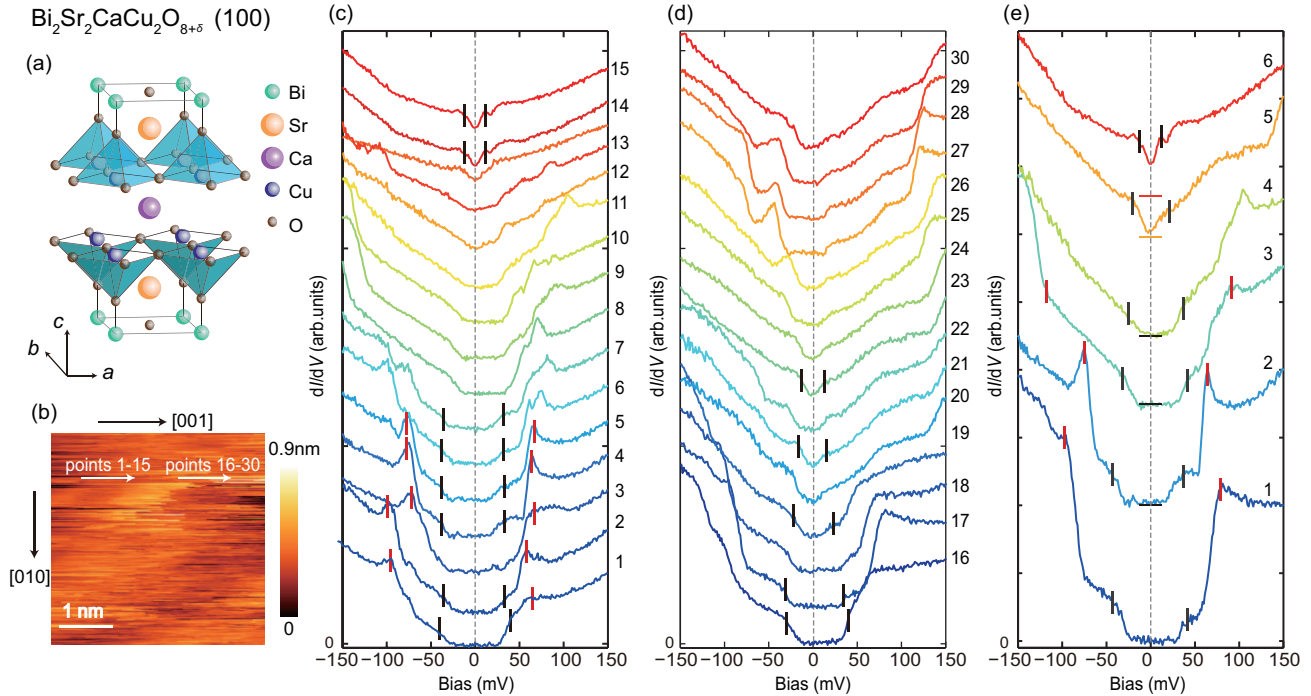}
%}
\caption{\label{fig2}(Color online).  (a) Cartoon of half a unit cell in the Bi2212 crystal structure. (b) Topographic image ($V$=~1 V, $I$= 30~pA) of an area of 4 $\times$ 4 nm$^2$ on the cleaved Bi2212 (100) plane. The scan orientation is marked by black arrows. Note that since the image is not atomically resolved, there might be a slight angular misalignment from the marked orientation.  (c,d) The STS spectra (setpoint: $V$= 100~mV, $I$= 150~pA, $\Delta$$V$= 2~mV) taken along the cuts marked in (b), offset vertically for clarity. (e) Representative STS spectra. Horizontal bars indicate the zero position of dI/dV. Black vertical lines indicate the gap edge or coherence peaks. All of the spectra (including the spectra shown in Figs.~3 and 4) are measured at 4.5~K.}
\end{figure*}
%\medskip
%%%%%%%%%% Fig. 2 %%%%%%%%%%%

Experiments were conducted on Bi2212 single crystals and YBCO films in a Createc cryogenic STM system at 4.5~K. The single-crystalline 600~nm-thick YBCO  film on MgO was purchased from Theva. As shown in Figs.\ 1(b,c), both the Bi2212 and YBCO samples show a superconducting transition temperature ($T_c$) of 88~K. Fig.\ 1(a) shows a schematic of our experimental setup for cross-sectional tunneling. The Bi2212 crystal was fixed on a silicon substrate, and then mounted vertically on the holder. A notch was usually made on the silicon substrate to facilitate cleaving at 80 K to expose a side surface.  We also measured the side surface of the YBCO film grown on MgO in a similar way. Because MgO is insulating, silver epoxy with a thickness of tens of microns was placed on the film before mounting. It cleaves together with the film and was used for landing the STM tip. Electrochemically-etched tungsten tips were used for all STM measurements after careful treatment on a Au (111) surface. The dI/dV spectra were collected using a standard lock-in technique with modulation frequency $f = 975$~Hz.

%%%%%%%  Fig.3 %%%%%%%%%
%\vskip 0.5\baselineskip
%\vskip 4mm
\begin{figure*}
%\centerline{
\includegraphics[width=\textwidth]{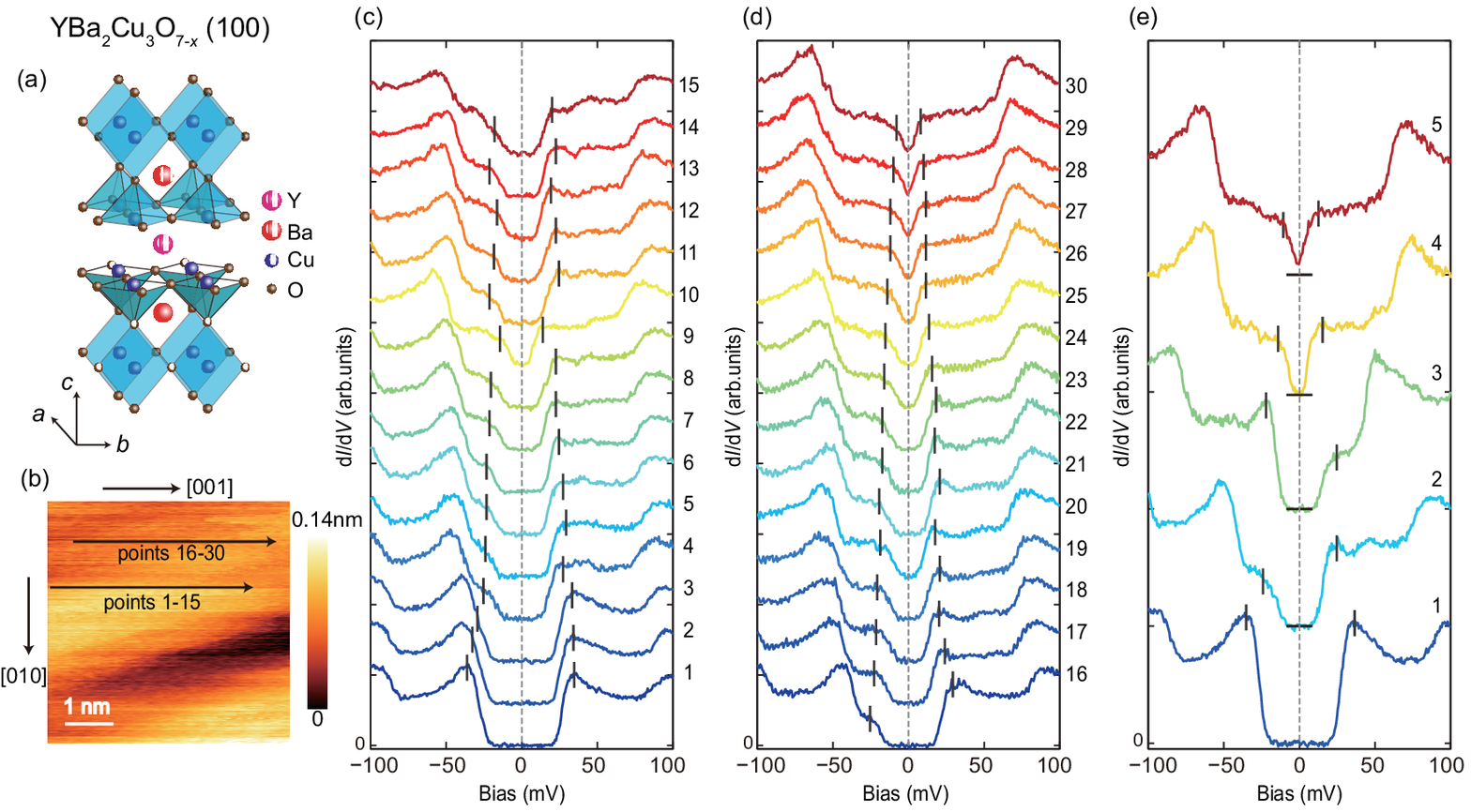}
%}
\caption{\label{fig3}(Color online). (a) Cartoon of the YBCO crystal structure. (b) Topographic image ($V$= -70 mV, $I$= 100 pA) of cleaved YBCO (100) surface of size of 5 $\times$ 5 nm$^2$, scan orientation as marked.  (c,d) STS spectra (setpoint: $V$= 80 mV, $I$= 100 pA, $\Delta$$V$= 2 mV) taken along the [001] direction with spatial spacing of 0.5 nm.  (e) Representative STS spectra. The horizontal bars indicate the zero position of dI/dV, and black vertical lines mark the coherence peaks or gap edge.}
\end{figure*}
%\medskip
%%%%%%%  Fig.3 %%%%%%%%%

%%%%%%%  Fig.4 %%%%%%%%%
%\vskip 0.5\baselineskip
%\vskip 4mm
 \begin{figure*}
%\centerline{
\includegraphics[width=\textwidth]{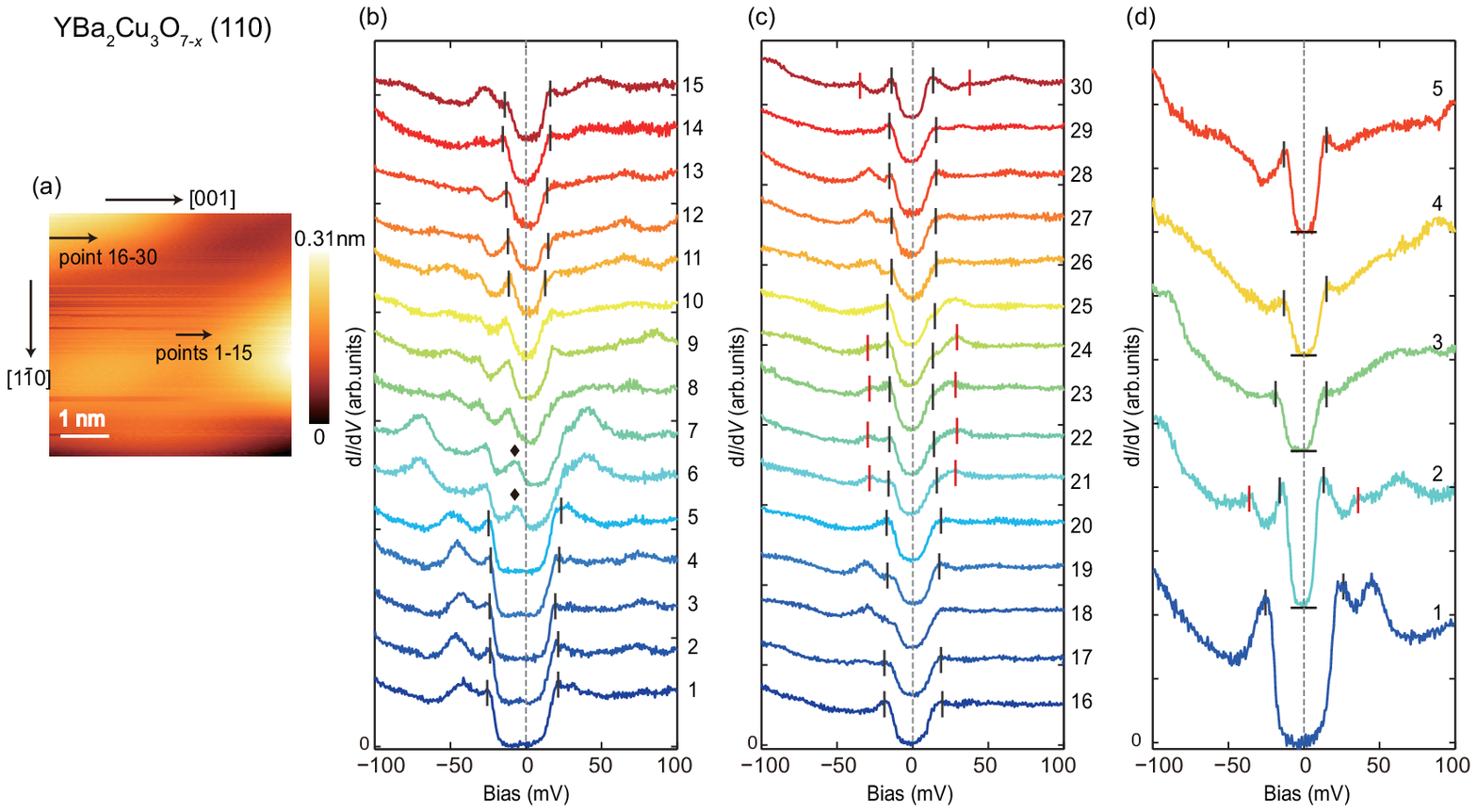}
%}
\caption{\label{fig4}(color online) (a) Topographic image ($V$= 100 mV, $I$= 100 pA) of a cleaved YBCO (110) surface of size of 5 $\times$ 5 nm$^2$,  scan orientation as marked. (b,c) STS spectra (setpoint: $V$= 100 mV, $I$= 100 pA, $\Delta$$V$= 2 mV) taken along the arrows marked in (a).  (d) Representative STS spectra.  The horizontal bars indicate the zero position of dI/dV. Black vertical lines mark the coherence peaks, red vertical lines indicate the possible second gap or bosonic mode, and diamonds indicate possible in-gap peaks.}
\end{figure*}
%\medskip
%%%%%%%  Fig.4 %%%%%%%%%

As illustrated in Fig.\ 2(a), Bi2212 is composed of periodically-stacked superconducting CuO$_2$ layers and charge reservoir layers (BiO, SrO), with a half $c$-axis lattice constant of 1.54~nm between adjacent copper oxide bilayers. Local tunneling to the (100) plane enables direct measurement of these different layers. Fig.~2(b) is a topographic image taken on the cleaved (100) surface of Bi2212.  Although the topography lacks atomic resolution, the overall roughness is less than 1~nm, and the tilt angle of this local surface to the ideal (100) plane is less than 7\,$^\circ$. Figs.~2(c,d) show two STS line-cuts taken over a distance of 1~nm along the $c$ direction, as marked by the arrows in Fig.~2(b), and show strong spatial dependence. Spectrum 1 in Fig.~2(c) shows a large gap-like structure reaching zero conductance, with a flat bottom over $\pm$25~meV, which evolves into spectrum 15 with just a weak V-shaped gap of 14~meV. Similar variation can also be observed in Fig.~2(d). Note that such a significant change happens on the length scale of 1~nm, a thickness corresponding to several atomic layers and less than the spacing between adjacent CuO$_2$ bilayers. Therefore, it is most likely that the different types of spectrum are mainly contributed by tunneling into different atomic layers. In Fig.~2(e), we plot several representative spectra selected from over 200 measured spectra. Nearly two-thirds of our spectra, as represented by curves 1-4 in Fig. 2(e), show a flat bottom, ranging in width from  $\pm$10 meV to $\pm$25 meV. The gap size defined by the shoulders outside the flat bottom, as indicated by the black vertical lines, is in the range of 20$\sim$40~meV. Many of these spectra additionally show nearly symmetric shoulders or peaks at higher energies of 60$\sim$90~meV (short red lines), which could be indications of an even larger gap. Other spectra exhibit V-shaped gaps, as represented by curves 5 and 6 in Fig.~2(e), with finite DOS at zero bias. Some have coherence peak-like features, as marked by the black vertical lines. The sizes of V-shaped gap are typically 13$\sim$18~meV.

We note that several previous cross-sectional STM studies of Bi2212 crystals have observed a V-shaped gap or a gap with a rounded bottom, but did not find U-shaped gap structure with a flat zero-DOS bottom$^{[14, 15]}$. The discrepancy may be due to differences in sample preparation. Our samples were cleaved in vacuum, while Refs.~[14] and [15] prepared the samples with diamond-filing or a razor blade outside the vacuum chamber.

%YBCO(100)
To establish whether the remarkable spatial dependence is specific to Bi2212, we performed similar measurements on YBCO thin films. Unlike Bi2212, the YBCO crystal structure, illustrated in Fig.~3(a), contains CuO chains which are known to be metallic. Fig.~3(b) shows a topographic image taken on the cleaved YBCO(100) surface, with overall roughness less than 0.14~nm. Figs.\ 3(c,d) show two STS line-cuts taken over 4.2~nm along the c-axis direction, as marked in Fig.~3(b). One can see a well-defined U-shaped gap with a flat bottom and coherence peaks in Fig.~3(c), which gradually evolves into a small V-shaped gap over several nanometers. Similar evolution can also be observed in Fig.~3(d). In Fig.~3(e) we show several representative spectra selected from over 500 spectra measured on the YBCO(100) plane. The typical U-shaped gap, as represented by curve 1, extends over $\pm$35~meV as defined by the coherence peaks (marked by black vertical lines). Smaller gaps with flat bottoms but asymmetric coherence peaks can also be observed (curves 2-4), with the gap amplitude (at either side of zero bias) ranging from 10 to 35~meV. The amplitude of the V-shaped gap such as that in curve 5 is only $\sim$8~meV, with non-zero DOS at zero bias.

%YBCO(110)
A U-shaped gap usually indicates fully-gapped superconductivity. However, the tunneling spectrum of an anisotropic superconductor may depend on the tunneling direction. This is because electrons with velocity perpendicular to the surface will have larger tunneling probability than  the others.$^{[16]}$. To check the possible anisotropy of the gap, we further measured STS on YBCO films cleaved along the (110) plane. Fig.~4(a) is a topographic image taken on the cleaved YBCO(110) surface, with overall roughness less than 0.31~nm. As shown in Figs.\ 4(b,c), STS line-cuts taken on the (110) surface (over a distance of 1~nm along c-axis) also display a U-shaped gap with clear coherence peaks. The gap also varies in size spatially, but less significantly than on the (100) plane. Sometimes peaks inside the gap but away from zero bias can be observed (marked by diamonds in Fig.4(b)), which need further investigation. Fig.~4(d) shows several typical spectra selected from more than 500 spectra, in which the gap size varies from 10 to 25~meV at different locations (coherence peaks are marked by black vertical lines). The maximum gap size on the (110) plane is reduced compared with the (100) plane, suggesting a narrower gap in YBCO along the [110] direction than the [100] direction. We note that some spectra display a second peak at higher energy, as marked by the red vertical lines in Fig. 4(c). These could indicate a double superconducting gap or originate from bosonic modes.
 
%\vskip 0.5\baselineskip
%\vskip 4mm
\begin{figure*}
 %\centerline{
 \includegraphics[width=\textwidth]{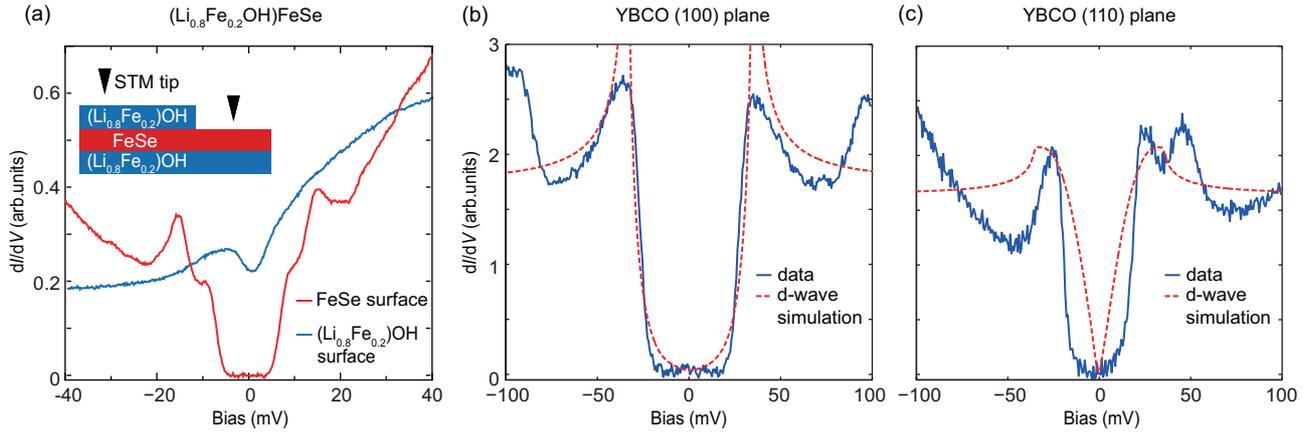}
 %}
\caption{\label{fig5}(color online) (a) Tunneling spectra on (Li$_{0.8}$Fe$_{0.2}$OH)FeSe taken from Ref.~[21]. The red and blue curves are taken on an FeSe-terminated surface and a (Li$_{0.8}$Fe$_{0.2}$OH)-terminated surface, respectively, displaying different gap structures. The inset sketches the two surface terminations. (b, c) The calculated tunneling spectra (dashed curves) to simulate the representative spectra (solid curves) taken on the YBCO (100) plane (b), and (110) plane (c), respectively. The simulations adopt a $d$-wave gap. Details are described in the text. }
\end{figure*}
%\medskip

%  discussion part
There are  certainly many unknown  issues to be further investigated for such rare cross-sectional STM measurements on cuprates. For example, since we have not obtained atomically-resolved cross-sectional images, the surface condition, such as reconstruction and defects, is not clear. 
However, the observed well-defined lineshapes of the differential conductance, together with their systematic evolution with  high spatial sensitivity, indicate that certain relevant information can still be retrieved from the data. One important finding is that the low-energy electronic structure of both Bi2212 and YBCO is strongly spatially dependent. 
The data thus force us to question the prevailing wisdom that the low-energy electronic states of cuprate superconductors are solely contributed by the CuO$_2$ planes.  Furthermore, we find that there are generally two types of gaps ---  the large gap with coherence peaks around extended U-shaped fully-gapped regions, and the small V-shaped gap which often has finite DOS at zero bias. The origins of such an inhomogeneity and particularly the U-shaped large gap require further study. For example,  the effect of  tunneling matrix elements may need to be considered, and non-local effects may play a role $^{[17]}$; the consequences of various competing phases in cuprates, such as charge density wave and pair density wave with $s$-wave form factors, should be examined $^{[18-20]}$;
and there is even the possibility of a small insulating gap in the charge reservoir layer. Moreover, if the V-shaped gap is the usual $d$-wave gap, one should investigate why its amplitude is anomalously small compared with previous studies on the (001) plane.

Alternatively, the observed well-defined gap structures and their spatial evolution, particularly those in YBCO, suggest another plausible explanation. Different layers  may have drastically different superconducting gaps, and the small V-shaped gap is induced by the large U-shaped gap situated in the CuO$_2$ planes through proximity effects. That is, the previous ARPES and STS data obtained through the $ab$ plane may be layer-integrated. 
Fig.~5(a) presents the realization of such a scenario in (Li$_{0.8}$Fe$_{0.2}$OH)FeSe, an iron-based superconductor with a $T_c$ of 40~K.$^{[21]}$ In this case, We observed a U-shaped nodeless 15~meV gap on the FeSe plane, and a small V-shaped (3$\sim$4~meV) gap with finite DOS at zero bias on the Li$_{0.8}$Fe$_{0.2}$OH surface, which is metallic due to surface charge reconstruction. The small gap is most likely due to the proximity effects from the superconducting FeSe layer. Therefore, the observed V-shaped gap in Bi2212 and YBCO can be naturally explained by similar proximity effects from the CuO$_2$ planes into the charge reservoir layers or CuO chains.

% simulation
The $d$-wave pairing symmetry in the cuprates has been established based on extensive experimental data. One critical signature of the $d$-wave gap is its nodal gap structure, which gives a finite DOS at the Fermi energy to the lowest temperatures. This has been demonstrated by ARPES$^{[6]}$, STM$^{[7]}$, and microwave superfluid density measurements$^{[22]}$, and by many other techniques.
However, the observation of a U-shaped gap on both YBCO (100) and (110) surfaces is apparently hard to square with a nodal $d$-wave pairing. To have a qualitative understanding, we simulated the tunneling spectra on the (100) and (110) planes by following the empirical model based on the WKB approximation $^{[15]}$.   The anisotropic band dispersion in the CuO$_2$ plane is found to be less important for cross-sectional tunneling $^{[15]}$, thus we assume an isotropic  dispersion  for simplification.
The gap function is taken as $\Delta(\theta) = \Delta_0  \cos(2\theta)$ for a $d_{x^2-y^2}$ symmetry ($\theta = 0$ is defined as the [100] direction), which gives low energy DOS of $\rho_0 (E,\theta) = Re[ E/ {\sqrt[]{E^2-\Delta(\theta)^2}}]$. The k-dependence of the tunneling probability is considered and is  given by the phenomenological factor$^{[16]}$: $p(\theta) = \exp\left[-\beta \sin^2(\theta-\theta_0)\right]$, where $\theta_0$ is defined by the surface normal. $\beta$ depends on $E_F$ and the details of the tunneling barrier. The tunneling conductance is then given by:
\begin{eqnarray*}
\frac{dI}{dV}\propto\int p(\theta) \rho_0 (E,\theta) \frac{df(E+eV)}{dV} dE d\theta
\end{eqnarray*}
where $f(E)$ is the Fermi distribution function.  In Fig.~5(b,c), we show the simulated spectra together with typical measured gaps on the YBCO (100) and (110) planes, respectively.  We chose $\Delta_0$ = 35~meV and $\beta$=8 (following Ref.~[15]). For the (100) case, the $d$-wave gap can produce a U-shaped spectrum, because the $p (\theta)$ factor leads to a greater contribution from states in the antinodal region. However, the simulated gap still has a rounded bottom, which differs from the measured U-shaped gap with an extended flat bottom. For the (110) case, the simulated gap is clearly V-shaped, as the nodal region dominates. If allowing a weak $k$-dependence of the $p (\theta)$ factor (i.e. small $\beta$), the gap on the (100) and (110) planes will be both V-shaped. Therefore, simulations based on a $d$-wave gap cannot readily reproduce the measured U-shaped gap.

Another critical consequence of the $d_{x^2-y^2}$ pairing symmetry is a surface Andreev bound state at zero energy when tunneling through the (110) plane $^{[23]}$, due to the phase change around gap nodes. A zero bias peak has been previously observed on the surface of [110]-oriented YBCO thin films with an inert surface prepared \textit{ex situ} $^{[24]}$. However, we did not observe such a peak at zero bias on cleaved YBCO along the (110) plane (Fig.~4), although the local surface condition needs to be further examined to check whether it  suppresses the bound state.

Our findings thus pose challenges for the current understanding of the cuprate superconductors. Further investigation is needed to reconcile with the existing evidence for $d$-wave superconductivity. For example, trapped flux in the so-called tri-crystal experiment$^{[25]}$ or the phase shift in DC SQUIDs,$^{[26]}$ which are sensitive to the phase of the superconducting order parameter and are considered smoking-gun evidence for a $d$-wave gap. On the other hand, our data are consistent with existing evidence for isotope effects,$^{[27]}$ and the electron-phonon interaction,$^{[28]}$ which support extended-$s$-wave pairing scenarios.$^{[29]}$

% discuss pseudo gap ?  no direct evidence, no discussion
% In the underdoped regime, anisotropic $d$-wave-like pseudogap has been observed, whose origin 

%  Summary 
In summary, our cross-sectional STS data on Bi2212 and YBCO show that their electronic structures have strong spatial dependence along the layer-stacking direction.  We observe a possible nodeless and anisotropic superconducting gap on both (100) and (110) planes that cannot be readily explained by the $d$-wave pairing symmetry. Our data provide a new dimension of information for depicting a more comprehensive picture of  cuprate superconductors. Further experimental and theoretical investigations are needed to fully understand these extraordinary findings.

\section*{\Large\bf References}%
%\begin{reference}
\vspace*{-0.8\baselineskip}

\hskip 7pt {\footnotesize%

%\bibliographystyle{unsrt}
%\bibliography{refs}

%\bibitem{cu1}
\REF{[1]} Bednorz J G and Mueller K A 1986 {\it Z. Phys. B} \textbf{64} (2) 189-193
\REF{[2]} Uchida S I   2015 {\it High temperature superconductivity, the road to high critical temperature} (New York:Springer)

%band calculations
 \REF{[3]}Massidda S, Yu J and Freeman A J 1988,  \textit{Physica} C \textbf{152} 251 
 \REF{[4]}Pickett W E  1989 \textit{Rev. Mod. Phys.} \textbf{61} (2) 433-512

%ARPES
\REF{[5]} Damascelli A, Hussain Z and Shen Z X 2003 \textit{Rev. Mod. Phys.} \textbf{75} 473
\REF{[6]} Shen Z-X et al 1993 {\it Phys. Rev. Lett.} \textbf{70} (10) 1553-1556
%STM
\REF{[7]} Renner Ch and  Fischer O 1995 {\it Phys. Rev.} B \textbf{51} (14) 9208-9218
\REF{[8]} Pan S H et al 2000 {\it Nature}  \textbf{403} 746-750
\REF{[9]} Fischer O  et al 2007 \textit{Rev. Mod. Phys.} \textbf{79} (1) 353-419

% Xue new
\REF{[10]}  Zhong Y et al  2016 \textit{Sci. Bull.}  \textbf{61} (16) 1239-1247
\REF{[11]}  Lv Y F et al  2015   {\it Phys. Rev. Lett.}  \textbf{115} 237002
\REF{[12]} Lv Y F et al  2016 {\it Phys. Rev.} B \textbf{93}  140504

% c-axis tunneling explains CuO2 plane tunneling 
\REF{[13]}  Misra S, et al 2002   {\it Phys. Rev. Lett.}  \textbf{89} (8) 087002

% Previous Cross sectional STM

\REF{[14]}  Hasegawa T, Kitazawa K, 1990 {\it Jpn. J. Appl. Phys.}  \textbf{29} L434
\REF{[15]}  Suzuki K  et al  1999   {\it Phys. Rev. Lett.}  \textbf{83} (3) 616-619

% tunneling theory 
\REF{[16]} E. L. Wolf, 1989 {\it Principles of Electron Tunneling Spectroscopy} (New York: Oxford University Press)

%Origin of the Electron-Hole Asymmetry in the Scanning Tunneling Spectrum of the High-Temperature Bi2Sr2CaCu2O8þ Superconductor
\REF{[17]} Nieminen J et al 2009 {\it Phys. Rev. Lett.} \textbf{102}  037001

% Raman-Scattering Measurements and Theory of the Energy-Momentum Spectrum for Underdoped Bi2Sr2CaCuO8þ Superconductors: Evidence of an s-Wave Structure for the Pseudogap
\REF{[18]} Sakai S et al 2013 {\it Phys. Rev. Lett.} \textbf{111}  107001

%Detection of a Cooper-pair density wave in Bi2Sr2CaCu2O8+x
\REF{[19]} Hamidian M H et al 2016 {\it Nature}  \textbf{532} 343-347

%Bidirectional Density Waves Cause Sub-Gap Structures but no Pseudogap in Superconducting Cuprates
\REF{[20]} Verret S, Charlebois M, S\'en\'echal D, and Tremblay A M S, arXiv:1610.01109 (unpublished).

%ours LiFeOHFeSe
\REF{[21]}   Yan Y J  et al  2016 {\it Phys. Rev.} B \textbf{94}   134502  

% microwave measurements
\REF{[22]} Hardy W N et al 1993 {\it Phys. Rev. Lett.} \textbf{70} (25) 3999-4002

% Andreev bound state
\REF{[23]} Hu C R, 1994 {\it Phys. Rev. Lett.} \textbf{72} 1526
\REF{[24]}  Yeh N C et al 2001 {\it Phys. Rev. Lett.} \textbf{87} (8) 087003

% phase sensitive
\REF{[25]} Tsuei C C et al 1994 {\it Phys. Rev. Lett.} \textbf{73} (4) 593-596
\REF{[26]} Wollman D A et al 1993 {\it Phys. Rev. Lett.} \textbf{71} (13) 2134-2137 

%isotope effects
\REF{[27]} Khasanov R et al 2004 {\it Phys. Rev. Lett.} \textbf{92} (5) 057602   
 
 %electron phonon interaction.
 \REF{[28]}Lanzara A et al 2001 {\it Nature}  \textbf{412} 510

% s-wave theory Chakravarty.
\REF{[29]}Chakravarty S et al  1993 {\it Science} \textbf{261}  337-340

}

%\end{reference}

\end{multicols}
\end{document}